\documentclass[10pt]{iopart}
\usepackage{graphicx}

\begin{document}

\paper[New results for the $t-J$ model in ladders]{New results for the $t-J$ model in ladders: 
Changes in the spin liquid state with applied magnetic field. Implications for the cuprates.}

\author{A F Albuquerque\dag  and G B Martins\ddag}

\address{\dag\ Departamento de F\'{\i}sica, ICEX, Universidade Federal de Minas Gerais, Belo Horizonte, Brasil}

\address{\ddag\ Department of Physics, Oakland University, Rochester, MI 48309-4487}

\ead{martins@oakland.edu}

\begin{abstract}
Exact Diagonalization calculations are presented for the $t-J$ model in the 
presence of a uniform magnetic field. Results for $2\times L$ ladders ($L=8,10,12$) and $4 \times 4$
square clusters with 1 and 2 holes indicate that the diamagnetic response to a 
perpendicular magnetic field tends to induce a spin liquid state in the spin background. 
The zero-field spin liquid state of a two-leg ladder is reinforced by the magnetic
field: a considerable increase of rung antiferromagnetic correlations is observed for
$J/t$ up to $0.6$, for 1 and 2 holes. 
Pair-breaking is also clearly observed in the ladders 
and seems to be associated in part with changes promoted by the field in the spin correlations
around the
zero-field pair.
In the $4 \times 4$ cluster, the numerical results seem to 
indicate that the field-induced spin liquid state competes with the zero-field 
antiferromagnetic short-range-order, 
the spin liquid state being favored by higher doping 
and smaller values of $J/t$. It is interesting to note that the field-effect 
can also be observed in a $2 \times 2$ plaquette with 1 and 2 holes. 
This opens up the possibility of gaining a qualitative understanding 
of the effect.

\end{abstract}


\section{Introduction}

The longstanding problem of understanding the dynamics of a charge carrier in 
an antiferromagnetic (AF) background \cite{Brinkman} is at the center of the High-T$_{\rm c}$ puzzle. For 
example, one of the important controversies still not settled is deciding if a hole 
injected into an AF Mott insulator will keep its integrity \cite{eder1} or dissociate \cite{anderson1}. 
These options imply two completely different ways of looking at the corresponding ground states: the first is a 
Fermi-liquid and the second is not. In the first case, 
does the spin-polaron concept \cite{eder1} capture all the basic ingredients of 
charge carrier dynamics in the cuprates? 
In the second, does the Resonating Valence Bond state \cite{anderson1}, and the 
associated idea of spin-charge separation,  can lead to 
an understanding of the phases 
at higher doping? In addition, recent numerical results for the Hubbard model \cite{Jarrel}, 
consistent with optical measurements \cite{optical}, have raised the possibility of 
a {\it partial} spin-charge separation. 
The lack of an exact solution, either analytic or numeric, to strongly correlated models 
for the cuprates, like 
the $t-J$ or Hubbard models, is at the origin of this and other controversies. 
It has been realized for some time already, that in addition to the technical difficulties associated 
with solving the most common model Hamiltonians for the cuprates, the fact that all these 
models seem to have a variety of competing ground states with different 
order parameters makes it even more difficult to obtain a clear picture. This multiplicity 
of competing ground states is reflected in the richness of the cuprates phase diagram and 
has given high visibility to theories which explore quantum phase transition (QPT) ideas \cite{vojta}.
In some cases, this competition seems to be resolved through the mutual coexistence of two 
different order parameters, as in the specific case of the stripes observed through neutron scattering 
measurements \cite{tranquada,riso} or the recurrent observation of AF order in the 
superconducting (SC) phase \cite{julien} through Nuclear Magnetic Resonance (NMR), 
and muon-spin-rotation 
($\mu$-SR). It is important to notice that parts of the various aspects 
revealed by experiments emerge in results obtained with different methods applied to 
slightly different Hamiltonians; however the overall picture 
never seems complete or coherent. Until a theoretical breakthrough occurs, with the appearance of 
some new analytical or numerical method which can finally settle these issues, researchers in this 
field will keep an attentive eye in new experimental results which can provide some clue about 
the true nature of the dynamics of charge carriers in an AF background. 
 
In this respect, a recent experimental development has renewed interest in this subject: 
neutron scattering \cite{lake,riso2}, Scanning Tunneling Microscopy 
(STM) \cite{hoffman}, and NMR \cite{mitrovic} 
experiments have shown that the application of an external magnetic field can change the relative 
presence of two of the most fundamental order parameters in the cuprates: AF 
and SC orders. The presence of 
the first being enhanced at the expense of the second. 
Apparently, extensive regions of AF order 
(centered in the vortex cores) were observed to develop as the strength of the 
magnetic field increases. This seems to be true 
for different materials (YBCO, LSCO and BSSCO) at different doping concentrations \cite{muon}.
Although a complete experimental picture has not yet emerged, the neutron scattering experiments 
in optimaly doped LSCO, for example, have indicated a substantial increase of inelastic 
scattering intensity at low energy transfers with increasing magnetic field \cite{riso2}. 
This seems to imply that the field induces slowly fluctuating AF spin correlations. However, 
several questions still remain: Is this effect generic to all cuprate families at all 
dopings? In what circumstances is the field-induced magnetism static 
and when is it dynamic? And very importantly, 
is the field-induced AF magnetism restricted to the vortex cores or is it present everywhere in 
the system?

SO(5) theories \cite{Zhang} had predicted the possibility
of antiferromagnetism inside the vortex cores in cuprates \cite{Arovas}. The qualitative idea is
that once superconductivity is suppressed, either by impurities or magnetic field, AF order
should be enhanced. In addition, after the above mentioned experiments, mean-field
calculations in effective Hubbard-like Hamiltonians produced results in 
agreement with this qualitative idea \cite{mean-field}.
This however comes as no surprise, since these effective Hamiltonians already include explicitly a
term for pairing and an on-site Coulomb repulsion, which favors AF order. 
For these mean-field calculations, one expects that
the suppression of one order will automatically lead to the enhancement of the other. 
QPT theories, on the other hand, state that the above experiments 
are indication that the superconducting phase is in the vicinity of a bulk QPT to a state with 
microscopic coexistence of superconducting and spin density wave orders \cite{Demler}.
It is also claimed that the STM results showing the appearance of charge density 
wave order can be accounted for by QPT-type theories \cite{Park}.

Two of the most popular models for the cuprates 
are the $t-J$ and Hubbard models. Much of what is known about how charge carriers move 
in an AF two-dimensional (2D) background was obtained through their
analysis. Prominent among the methods used for that are numerical techniques \cite{Elbio}. 
However, to the best of our knowledge, no non-mean-field calculations of a strongly 
correlated Hamiltonian (Hubbard- or $t-J$-like)
have been performed to analyze the effect of an external magnetic field upon the spin correlations. 
Previous numerical results for small square clusters were concentrated in the issues 
of spin-charge separation \cite{beran} and 
magnetic susceptibility of the ground state \cite{Veberic1}, but neither work analyzed the evolution of 
spin-spin correlations with the external magnetic field. Unfortunately, these two works called very little 
attention. One of the reasons may have been the fact that they were published before the main 
experimental results showing the effect of the magnetic field were performed. Another factor 
is the difficulty, mainly after introducing the magnetic field, of performing finite-size 
scaling analysis for square clusters, a necessary step in ascertaining any claims based on the 
numerical results \cite{Veberic1}. There is however an alternative route in 
obtaining credible  numerical results which 
can extend these two previous works, and provide some clues about the overall problem of charge 
carrier dynamics in an AF background: performing the calculations in $2 \times L$ ladders.
For some time already, 2-leg ladders have been a fruitful testing ground 
for the numerical study of strongly correlated models relevant to the cuprates \cite{elbio1}. 
This is especially true for exact diagonalization (ED) calculations, where the 
quasi-unidimensional aspect of ladders 
makes it feasible the study of finite-size effects, providing credibility to the results
 \cite{ladd-num}. Besides that, given the above mentioned difficulty in extracting uncontroversial 
conclusions 
from numerical results obtained for square clusters (with either ED or Density Matrix Renormalization
Group (DMRG) \cite{dmrg} methods), some of the more solid numerical results obtained for two-leg ladders 
have been qualitatively `translated' into the 2D cuprate arena \cite{scalapino,mart-pair}.

In this paper, the authors present zero-temperature ED calculations for the $t-J$ model with
a magnetic field in hole-doped 2-leg
ladders. All the calculations for ladders were done adopting periodic boundary conditions (PBC) 
along the legs.
The results show that the diamagnetic response of the system leads to a strengthening
of the rung spin correlations. This can be described in qualitative terms as a reinforcement
of the spin liquid (SL) character of the 2-leg ladder ground state.
In light of our conclusions for 2-leg ladders, and in keeping with the spirit of the
discussion in the previous paragraph, the
authors also performed calculations in small square clusters. The ED results for
$4 \times 4$ clusters presented intriguing results: the calculation
of spin correlations for different values of $J/t$, different dopings (1 and 2 holes), and using
different boundary conditions (PBC and open boundary conditions (OBC)) seem to indicate that
the movement of the holes, under the influence of the external field, promotes the
appearance of an SL state in the spin background. This is consistent with the
results for ladders; however, in the square clusters
this SL state has to compete with the zero-field AF short-range-order (SRO)
characteristic of doped cuprates.
Although the results are not conclusive \cite{newtjf}, they point to a possible new example of
competition
between different ground states in a strongly correlated 
model related to the cuprates \cite{vojta}. 
The rest of the paper is organized as follows. 
In the next section, a brief description of the Peierls transformation is given. Then, 
in section 3, results are presented for 2-leg ladders, followed by results for a $4 \times 4$ 
cluster. Summary, conclusions and prospects for future work are presented in the last section.

\section{Adding the field to the $t-J$ Hamiltonian: The Peierls transformation}

The magnetic field is introduced in the $t-J$ model through a Peierls transformation:
\begin{equation}
H = J \sum_{\langle {ij} \rangle} 
\lbrack{{{\bf S}_{i}}\cdot{{\bf S}_{j}}}-{{1}\over{4}}n_{i}n_{j}\rbrack
- t \sum_{\langle {ij} \rangle\sigma} \lbrack {c^\dagger_{i \sigma} c_{j \sigma} \exp(\rmi \theta_{ij}) + h.c.} \rbrack,
\end{equation}
where $c_{i \sigma}$ is a fermionic annihilation operator with double occupied states projected out, 
$\langle \  \rangle$ indicates that the summations include only nearest-neighbor (NN) sites, and 
$\sigma$ stands for the spin degree of freedom. The  
phase in the hopping term can be written, using a Landau gauge ${\bf A}=B(0,x,0)$, as 
\begin{equation}
\theta_{ij} = {{e}\over{h}}{{\bf r}_{ij}}\cdot{{\bf A({\bf r}_i})},
\end{equation}
and the magnetic field $B$ is related to the dimensionless flux per plaquette through $\alpha=2\pi Ba_0^2/\phi_0$, 
where $\phi_0$ is the unit flux quantum and $a_0$ is the lattice parameter \cite{note1}. 
It is important to remark that, as one of the main objectives of the present work is to 
gain insight into the dynamics of charge carriers introduced into an AF background, 
the calculations concentrate in the diamagnetic response of the charge carriers to the 
external field, 
and will therefore omit the Zeeman term. For laboratory-strength magnetic fields, this can be 
considered a reasonable approximation to the actual experimental situation. However, for the 
calculations presented here, mainly for square clusters, where the fields involved can be 
quite substantial \cite{flux}, the Zeeman term would have to be taken in account if one wishes to 
make any connection to the experimental results mentioned above
\cite{lake,hoffman,mitrovic}.

\section{Results for 2-leg ladders}
\subsection{Single-hole results}

As it is well known \cite{elbiorice}, the spin sector in ladders is gapped and therefore presents 
the exponential decay of AF spin correlations in the leg direction, characteristic of a SL state. 
In a qualitative sense, the SL character of the undoped ground state in ladders can be 
described as a collection of rung-singlets \cite{elbiorice}. This picture survives 
doping with holes up to a moderate concentration \cite{riera}. Therefore, to obtain 
a qualitative understanding of how the magnetic field affects the zero-field ground state, 
correlations between spins in the same rung (from now on called rung correlations) were 
calculated as a function of field. 
In figure 1a it is shown the evolution with field (in terms of $\alpha$) of the rung
correlations in a $2 \times 8$ ladder with 1 hole, for $J/t = 0.2, 0.4$ 
and $0.6$. It can be seen that there is a substantial increase of the rung correlations, while 
correlations for spins farther apart than $d = \sqrt{2}$ 
are always smaller than the zero-field value (see below). As the magnetic field influence 
on the spin background is being assessed through the 
hole movement, one would expect a more pronounced effect at lower values of $J/t$, when the hole has 
higher mobility and is more effective in driving the behavior of the spin background. 
As can be observed by comparing results for $\alpha=0$ and $\pi/4$, the increase of the spin 
correlations is larger for the smaller value of $J/t$. However, a substantial increase is still 
present for $J/t=0.6$.

\begin{figure}[h]
\centering
\begin{minipage}[c]{4.8cm}
\centering \includegraphics[width=5.0cm]{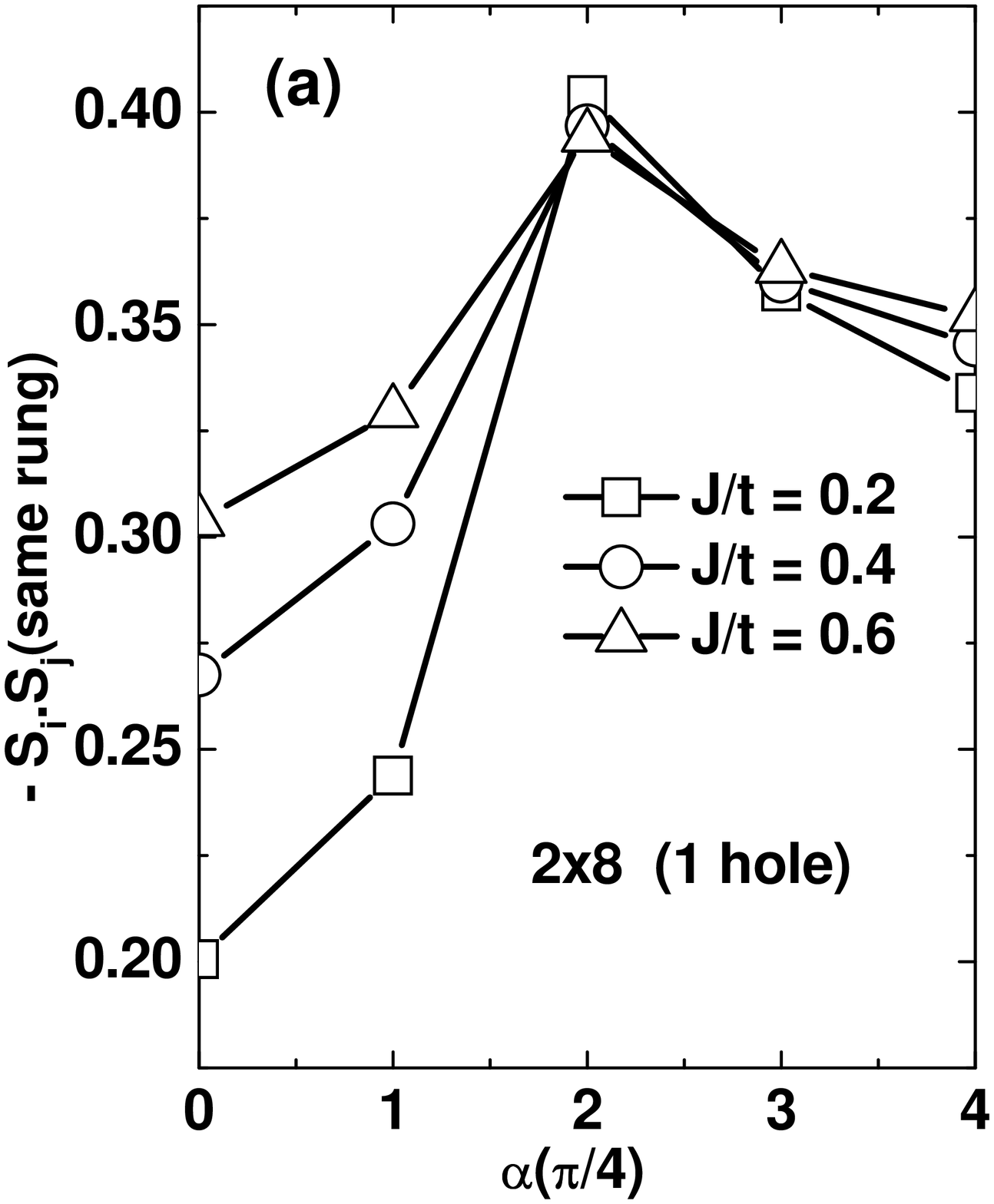}
\end{minipage}%
\begin{minipage}[c]{4.0cm}
\centering \includegraphics[width=6cm]{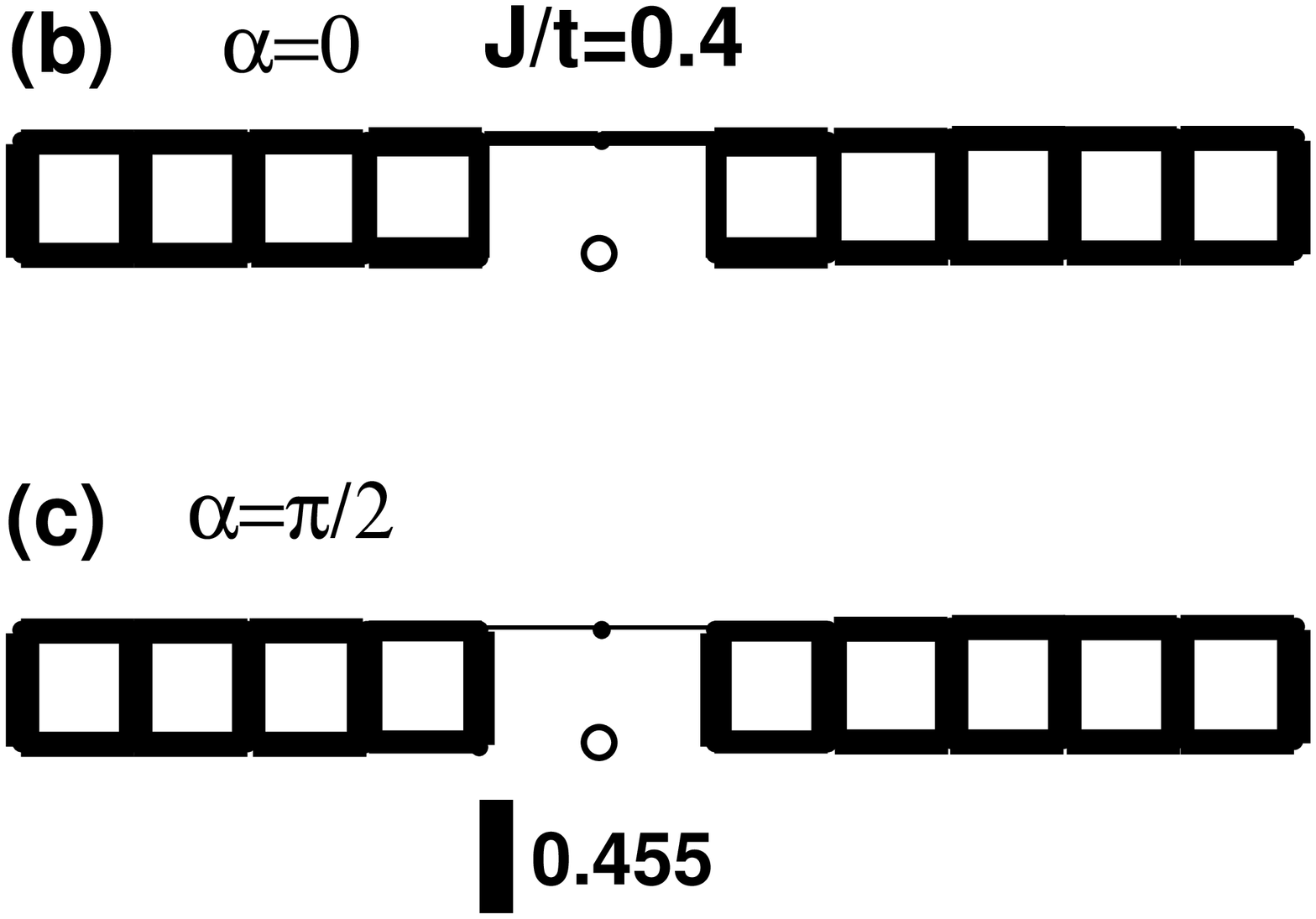}
\end{minipage}
\vspace{-0.5cm}
\caption{ (a) Variation with field of rung spin correlations for a $2
\times 8$ ladder with 1 hole, for $J/t=0.2, 0.4$ and $0.6$.
Note the substantial increase of the rung correlations as a function of magnetic
field, especially for lower values of $J/t$.
(b) and (c) 
$2 \times 12$ calculations for $J/t=0.4$ and 1 hole. Representation of the NN spin correlations 
after projecting the hole (open circle), for 
zero-field (b) and $\alpha=\pi/2$ (c). The thickness of the lines is proportional
to the absolute value of the correlations. All values are negative (AF).
In (b), note the weakening of the AF
spin correlations in the rungs close to the projected hole. In (c),
note how the rungs close to the hole have their AF spin correlations increased, 
compared to the zero-field case in (b), especially the ones adjacent
to the projected hole. The line at the bottom of (c) represents the rung correlation for 
an {\it undoped} $2 \times 12$ ladder.}
\label{fig1}
\end{figure}

It is interesting to take a closer look at how the 
spin correlations around the hole will change 
with the magnetic field. To do that, one can project out of the ground state wave function all the states 
where the hole occupies one specific site and and use these states to calculate the 
spin correlations. Figures 1b and 
1c compare the results for $\alpha=0$ and $\pi/2$ in a $2 \times 12$ ladder with 1 hole, for 
$J/t=0.4$. The thickness of the lines is proportional to the absolute value of 
the correlations (all correlations displayed are AF). 
For zero-field (figure 1b) one can see the distortion in the spin background caused 
by the presence of the hole (open circle in the figure): 
the rungs adjacent to the hole have their spin correlations substantially decreased, 
in comparison with rungs far away from it, which display correlations similar to the 
ones in the rungs of undoped ladders. For smaller 
values of $J/t$ the distortion of the spin background extends 
to larger distances, consistent with the higher mobility of the hole \cite{mart-dagg}. 
The application of the magnetic field, as can be seen in figure 1c, affects drastically the 
zero-field spin background: all 
rungs close to the projected hole have their AF spin correlations enhanced, 
particularly the ones adjacent to the hole, when compared to the zero-field values \cite{note-bonds}. 
One can describe the new spin arrangement
as a set of `strong rung-singlets'. As a reference, it is shown at the bottom of figure 1c a 
line which is representative of the rung correlations of an {\it undoped} $2 \times 12$ ladder, 
showing that some of the rung correlations for $\alpha=\pi/2$ are even larger than in the undoped system.
Similar calculations for smaller ladders ($L=8,10$) produce very similar results, indicating that 
any size effects present are small and do not affect the qualitative description of the results. 

Regarding size effects, one should note that, at zero-field for $J/t=0.4$ (figure 1b), rungs 
as far as 3 lattice spacings from the projected hole are affected, having their 
spin correlations {\it decreased} as compared to the undoped ladder. 
The fact that this differs very little from $2 \times 10$ results 
indicates that at zero-field the results are quite 
well converged for a $2 \times 12$ ladder. At finite-field (figure 1c), the 
distortion caused by the field, i.e., the {\it increase} in the rung correlations 
when compared to the zero-field results, has approximately the same reach 
(3 lattice spacings from the projected hole), indicating that the results 
seem well converged also with field for a $2 \times 12$ ladder. 
Now, results are presented for different ladder sizes. In figure 2a it is shown how the
rung correlations evolve for increasing ladder sizes ($L=8, 10$ and $12$) in the single-hole case.
Although the comparison
is not intended as a true finite-size-scaling analysis, since the hole doping decreases 
from $1/16$ to $1/24$, it indicates that the magnetic field effect does not seem
to vanish as $L$ increases. As discussed already in connection with figure 1c, the $J/t=0.4$
single-hole calculations in a $2 \times 12$ ladder are already representative
of qualitative bulk behavior at low doping. It would be interesting though to extend the
calculations here performed by using other methods (for example, DMRG) and also by performing
ED calculations with twisted boundary conditions \cite{tohyama}.

\begin{figure}[h]
\centering
\begin{minipage}[c]{4.7cm}
\centering \includegraphics[width=5.5cm]{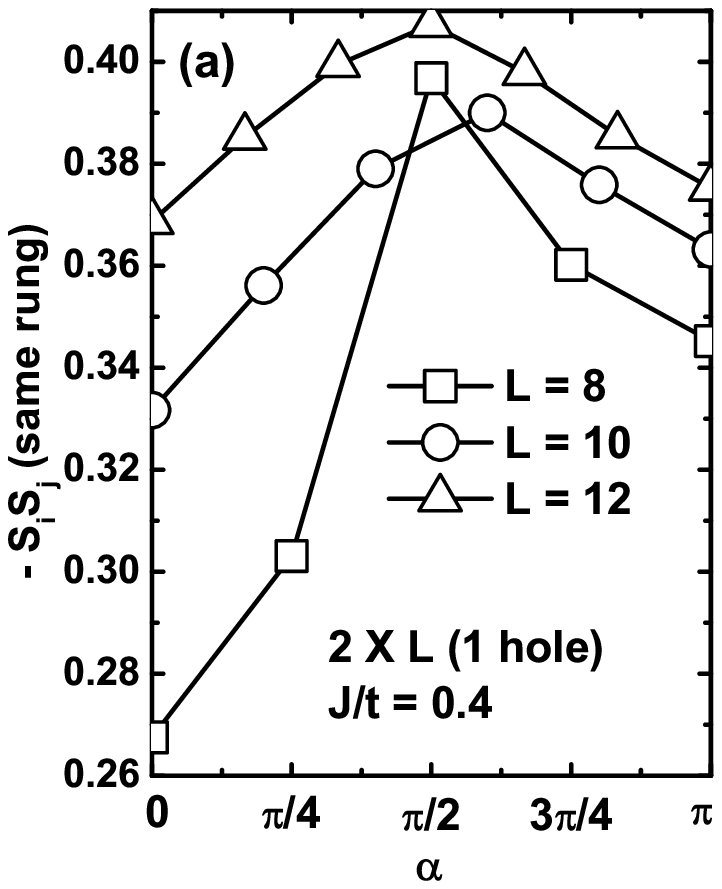}
\end{minipage}
\begin{minipage}[c]{4.7cm}
\centering \includegraphics[width=5.6cm]{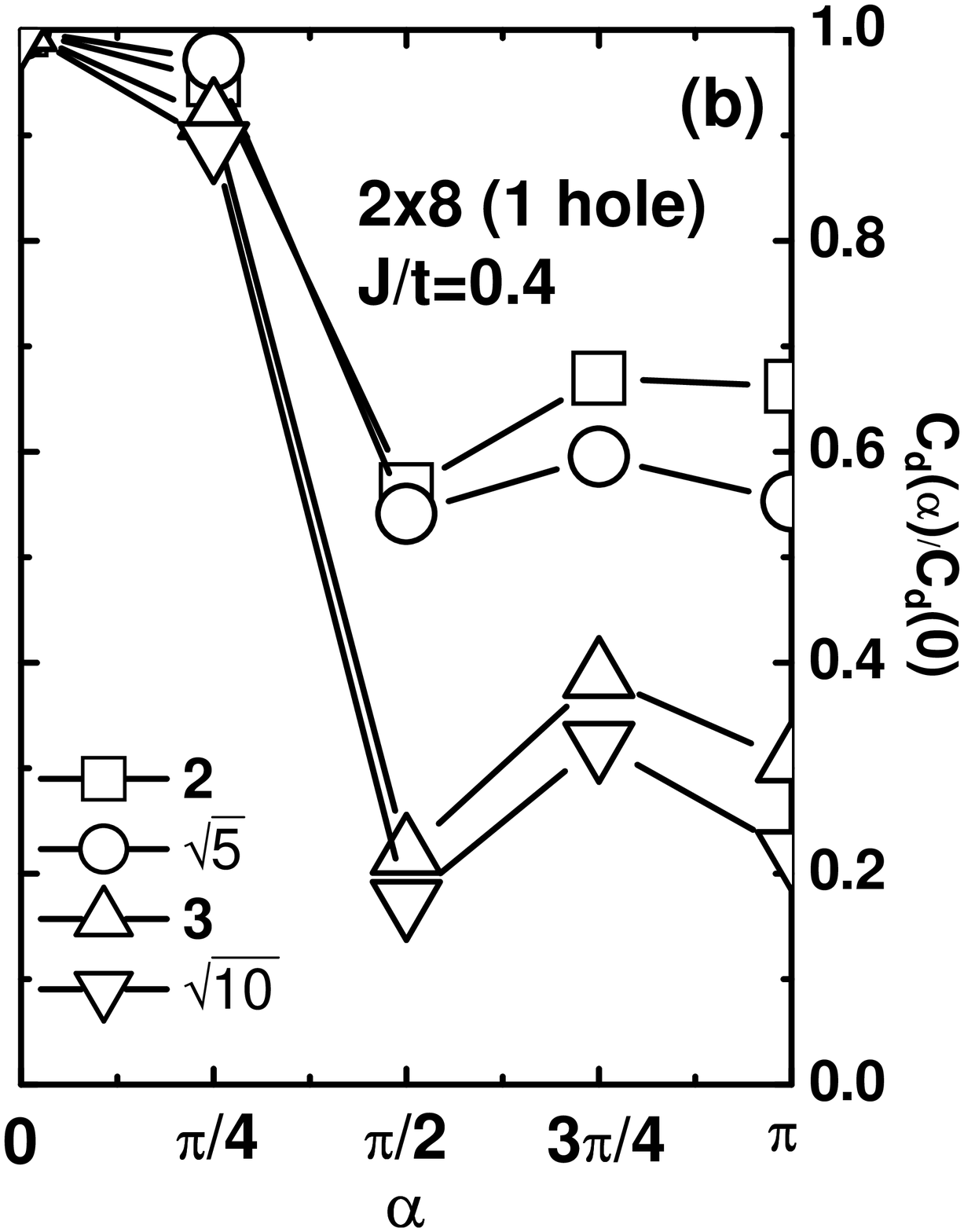}
\end{minipage}
\vspace{-0.5cm}
\caption{(a) Analysis of the evolution with magnetic field of the rung 
correlations for ladders of increasing length. 
Results are shown for $2 \times L$ ladders ($L=8,10,12$) with 1 hole, $J/t = 0.4$.
(b) Dependence with field of long-distance spin correlations in ladders: Ratio between
correlation at finite-field and zero-field for different distances. Note the sharp decrease
of the ratio (at all distances) as the field increases.
$C_d$ is defined as $\overrightarrow{S_i} \cdot \overrightarrow{S}_{i+d}$.
}
\label{fig2}
\end{figure}

To complete the analysis, calculations were done for longer range spin correlations.
One would expect that the increase of the rung correlations will lead to an overall 
decrease of the spin correlations along the legs (in the limit where the rungs form 
perfect singlets, the spins would be completely uncorrelated along the legs). 
In figure 2b, it is shown the field dependence of long-distance spin 
correlations in a $2 \times 8$ ladder with 1 hole for $J/t=0.4$.
The decrease with field of the ratio between finite- and zero-field correlations for 
all distances larger than 
$\sqrt{2}$ is consistent with an increase in the SL character of the ground state 
at finite-field \cite{sqrt2}.

\begin{figure}[h]
\centering
\includegraphics[width=10cm]{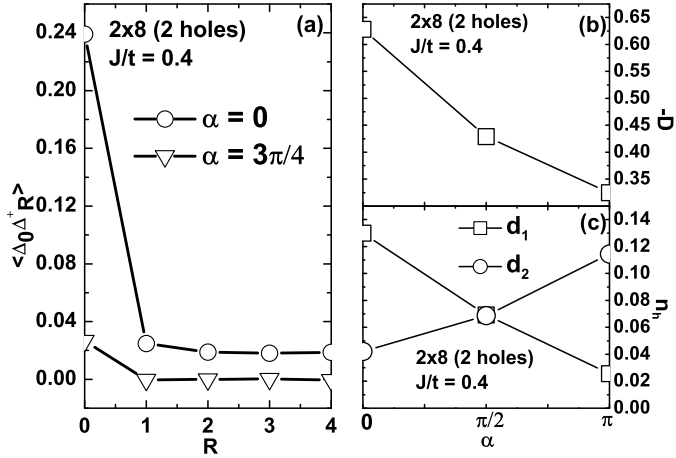}
\caption{(a) Pair-pair correlations for a $2 \times 8$ ladder with 2 holes for $J/t = 0.4$ 
at zero- (circles) and finite-field (squares). 
(b) Variation with field of spin correlations across a $\sqrt{2}$ diagonal 
when the two holes are projected in the 
other diagonal of the same plaquette (for same cluster and parameters as in (a)). 
(c) Field variation of the probabilities for two different hole configurations: holes in 
the diagonal of a plaquette ($d_1=\sqrt{2}$, squares) and holes at maximum distance 
($d_2=\sqrt{17}$, circles). 
The first is the most probable configuration at zero-field ($\alpha=0$) and the second 
is the most probable for $\alpha=\pi$ (same cluster and parameters as in (a) and (b)).
}
\label{fig3}
\end{figure}

\subsection{Adding a second hole: decrease of pairing correlations with applied field}

It is interesting to note that a second hole added to the system depicted in figure 1c 
will not tend to form a pair with the
first one. In reality, they tend to stay as far as possible from each other and 
a calculation of spin correlations shows that the `strong rung-singlets' 
picture still holds. In figure 3a, results are shown for pair-pair correlations \cite{pair-note} 
at zero- and finite-field 
for a $2 \times 8$ ladder with 2 holes and $J/t=0.4$. The results clearly show 
the pair-breaking effect of the magnetic field, as well as the loss of coherence 
in the pair propagation along the ladder. 
Our calculations also indicate that the presence of the magnetic field makes it 
energetically favorable for the pair to break and for the spin background to form 
strong rung-singlets, in contrast 
to zero-field, where the preferred configuration is the formation of a bound-pair, where the holes are 
separated by $\sqrt{2}$ with a strong {\it across-the-diagonal} AF 
spin correlation \cite{scalapino}. 
Figure 3b shows the variation with field of the across-the-diagonal spin 
correlation (denoted as {\bf D}) for a $2 \times 8$ ladder with 2 holes and $J/t = 0.4$. 
The suppression of the across-the-diagonal correlation 
by the field partially explains the pair-breaking effect.
Compared to $\alpha=0$, the value of {\bf D} for $\alpha=\pi$ has decreased in half. This 
is accompanied by an increase in the average separation of the two holes.
In figure 3c (again for $2 \times 8$, 2 holes, $J/t=0.4$), 
after one hole is projected at an arbitrary site, results are shown for the 
density $n_h$ of the second hole 
at two different distances from the projected hole.
The squares indicate results for distance $d_1=\sqrt{2}$ (most probable hole configuration 
at zero-field) and the circles for $d_2=\sqrt{17}$
 (most probable hole configuration at high-field).
At this point, the following observation is appropriate: at first sight, it appears 
counterintuitive that at finite-field the pair will be broken. 
After all, the single-hole results above have shown that the field has increased the overall 
strength of the rung correlations. And by using an 
argument based on the optimization of the ground state energy, it 
seems reasonable that the second hole
would pair-up with the first one in the same rung, thus minimizing the number of broken 
AF bonds in the rungs. However, as it was shown in the (zero-field) numerical work of White and
Scalapino \cite{scalapino}, the structure of the pair is more complicated. It is not only the 
magnetic energy that has to be optimized; the kinetic energy of the two holes has to be 
taken in account also: their ability to hop from a $\sqrt{2}$ configuration (when they are 
positioned in the diagonal of a plaquette) to a minimum distance configuration (positioned in 
one of the rungs of the same plaquette) is more important than just minimizing the number of 
broken AF bonds in the rungs. One evidence being the fact that (for values of $J/t$ 
around 0.4) the $\sqrt{2}$ configuration for the holes is more probable than a same- 
rung configuration. And, as found out by White and Scalapino \cite{scalapino}, the matrix 
element of the hopping term between these 
two configurations is critically dependent on the existence of a strong AF bond 
in one diagonal of the plaquette when the holes are in the $\sqrt{2}$ configuration, 
occupying the other diagonal. One has therefore the picture at zero-field of a 
pair of holes `straddling' the AF bond formed in one of the diagonals of a plaquette. Our results 
in figure 3b indicate that the presence of the field weakens considerably this `across-the-diagonal' 
spin configuration, leading to the `ungluing' of the pair (figure 3c).

\begin{figure}[h]
\centering
\includegraphics[width=10cm]{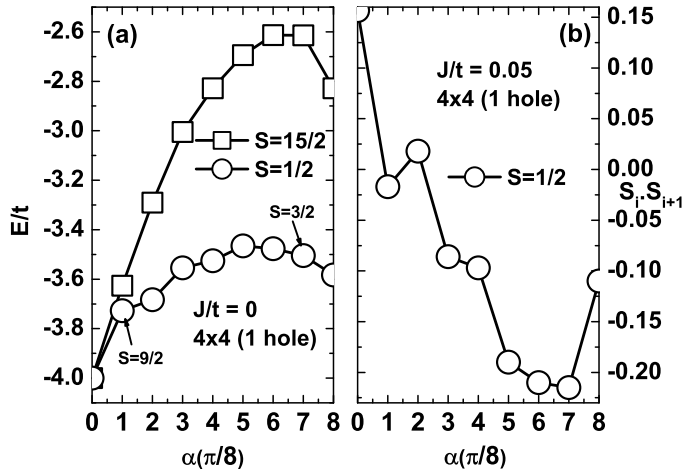}
\caption{(a) Variation of energy with magnetic field in a $4\times4$ cluster
with 1 hole for $J/t = 0$.
Squares: Ground state energy for $S=15/2$ sector. Circles: Absolute ground state energy. 
$S=1/2$ for all values of $\alpha$, with the exception of $\alpha=\pi/8$ ($S=9/2$) and $7\pi/8$
($S=3/2$).
(b) Variation of NN spin correlations with magnetic field for a $4 \times 4$ cluster 
with 1 hole for $J/t = 0.05$.
The field drives the system from a ferromagnetic state to a new state with AF NN spin 
correlations. 
}
\label{fig4}
\end{figure}

\section{Results for a $4 \times 4$ cluster}
\subsection{NN correlations for 1 and 2 holes}

Motivated by the interesting results obtained for ladders, the authors also performed calculations 
for the $t-J$ model in $4 \times 4$ clusters with 1 and 2 holes. As already mentioned above, 
extreme caution has to be exercised in interpreting the results. However, the authors believe 
the interpretation of the ladder results can help our intuition in 
understanding the $4 \times 4$ results. In an attempt to bolster the 
confidence in the results obtained, calculations were performed using PBC and OBC.
For reasons which will become clear soon, we start the calculations 
for the $4 \times 4$ cluster with values of $J/t < 0.1$.
In figure 4a, results for the energy variation with magnetic field are shown for $J/t = 0$ in 
a $4\times4$ cluster with 1 hole. The squares are results for the totally polarized $S=15/2$ sector
and the circles are results for the {\it absolute} ground state 
sector. For $J/t=0$, at zero-field, as it is well known, the ground state 
of the $t-J$ model with 1 hole is totally polarized
($S=15/2$, for this case) \cite{Nagaoka}. However, at low field (as seen in figure 4a), with just 
one flux quantum threading the cluster ($\alpha=\pi/8$), the polarization of the absolute 
ground state is already below saturation ($S=9/2$), and 
for all other values of $\alpha$ the total spin of the ground state 
is $S=1/2$, with the exception of 
$\alpha=7\pi/8$, where $S=3/2$. 
As noted in Veberi$\check{c}$ {\it et al.} \cite{Veberic2}, the field dependence of the 
energy for the $S=15/2$ sector (squares in figure 4a) is expected, as a 
consequence of the increase of the hole energy 
with the cyclotron frequency: $\delta E_h=eB/m_{eff}$ (however, see reference \cite{Hasegawa}). 
On the other hand, to the best of our knowledge, the change of the ground state sector from fully polarized 
at zero-field to $S=1/2$ at finite-field was never discussed.
It is natural to expect that, whatever the effect introduced by the magnetic field, the lowering 
of the total spin of the ground state to its minimum value should be associated 
with the development of AF correlations. 
To test this hypothesis, NN 
spin correlations in the $S=1/2$ sector were calculated for $J/t=0.05$ 
with 1 hole in a $4 \times 4$ cluster \cite{sizeeffect}. 
The results are presented in figure 4b, where it is shown that the magnetic field initially turns the 
ferromagnetic NN spin correlations into AF correlations and gradually enhances them. 
Similar results are obtained for $J/t=0$, indicating that 
it is not the gain in magnetic energy which drives the increase in the NN AF correlations. 
The results at higher values of $J/t$ are similar to the ones in figure 4b, 
{\it but only up to $J/t \approx {\it 0.07}$}. 
A small increase in the NN AF correlations can be achieved at higher values of 
field ($\alpha=7 \pi /8$) up to $J/t \approx 0.1$. Therefore, from the 
single-hole results at small values of 
$J/t$ it is clear that the field (when only the diamagnetic response is 
included in the Hamiltonian) is able to change radically the character of the ground state: 
from fully polarized to minimum polarization. However, this effect is gradually weakened, 
until it is no more present, as $J/t$ increases. It is well known from the early numerical 
studies and from photoemission measurements \cite{Elbio} that the hole bandwidth is 
strongly renormalized by $J$. As mentioned in the ladder discussions above, 
the size of the region (in the spin background) 
affected by the hole presence decreases as $J/t$ increases. In the limit where $J/t 
>> 1$, an injected hole would be unable to move \cite{phase-sep}. 
As the field effect is caused by the hole movement, it is natural 
to expect that it will become weaker as the ability of the 
hole to move is degraded by the stiffening of the spin background with 
increasing $J/t$. 
However, a value of $J/t=0.1$, although already introducing very short-range AF fluctuations 
in the spin background, still allows the hole a fair amount of freedom to move 
around the $4 \times 4$ cluster. Therefore, it is puzzling that the 
effect of the field in the NN correlations is already negligible for $J/t \approx 0.1$. 
(However, as is going to be shown later, the effect is more resilient at longer distances).
One possible explanation for the existence of a threshold 
at $J/t \approx 0.1$ in 
the field effect is that the hole movement under the influence of the 
field tries to induce a ground state with a {\it qualitatively different} spin background 
than the one at zero-field, and this sets in a competition between different ground 
states. For very small $J/t$, where the zero-field ground state is the Nagaoka phase \cite{Nagaoka}, 
the new `order' (imposed by the field and characterized by minimum polarization) 
quickly dominates. It is well 
known that the Nagaoka phase is very delicate and that small perturbations are 
able to drive the system out of it. Therefore, the effect in figure 4b 
is not hard to accept. However, for values of $J/t$ where the 
$4 \times 4$ cluster at zero-field has already moved out of the Nagaoka phase, 
to a more robust ground state, the field is not able to impose its new type 
of `order' to all distances. Therefore, if one could somehow `weaken' the zero-field 
ground state, the field-generated ground state would certainly be still 
visible at higher values of $J/t$. One way of weakening the zero-field 
ground state is to increase the hole concentration.

\begin{figure}[h]
\centering
\begin{minipage}[c]{5.0cm}
\centering \includegraphics[width=5.0cm,angle=0]{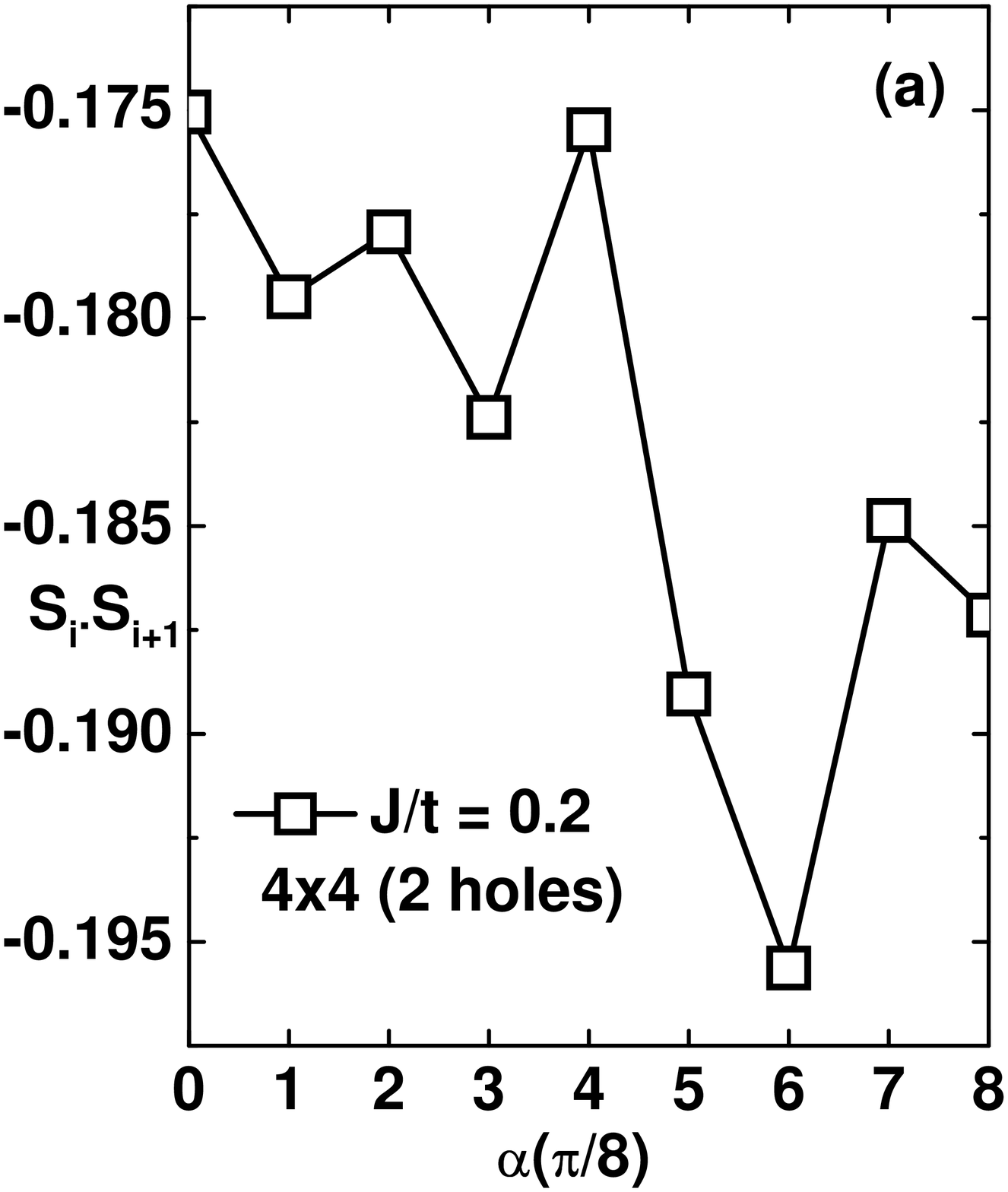}
\end{minipage}%
\begin{minipage}[c]{2.5cm}
\centering \includegraphics[width=2.5cm,origin=c]{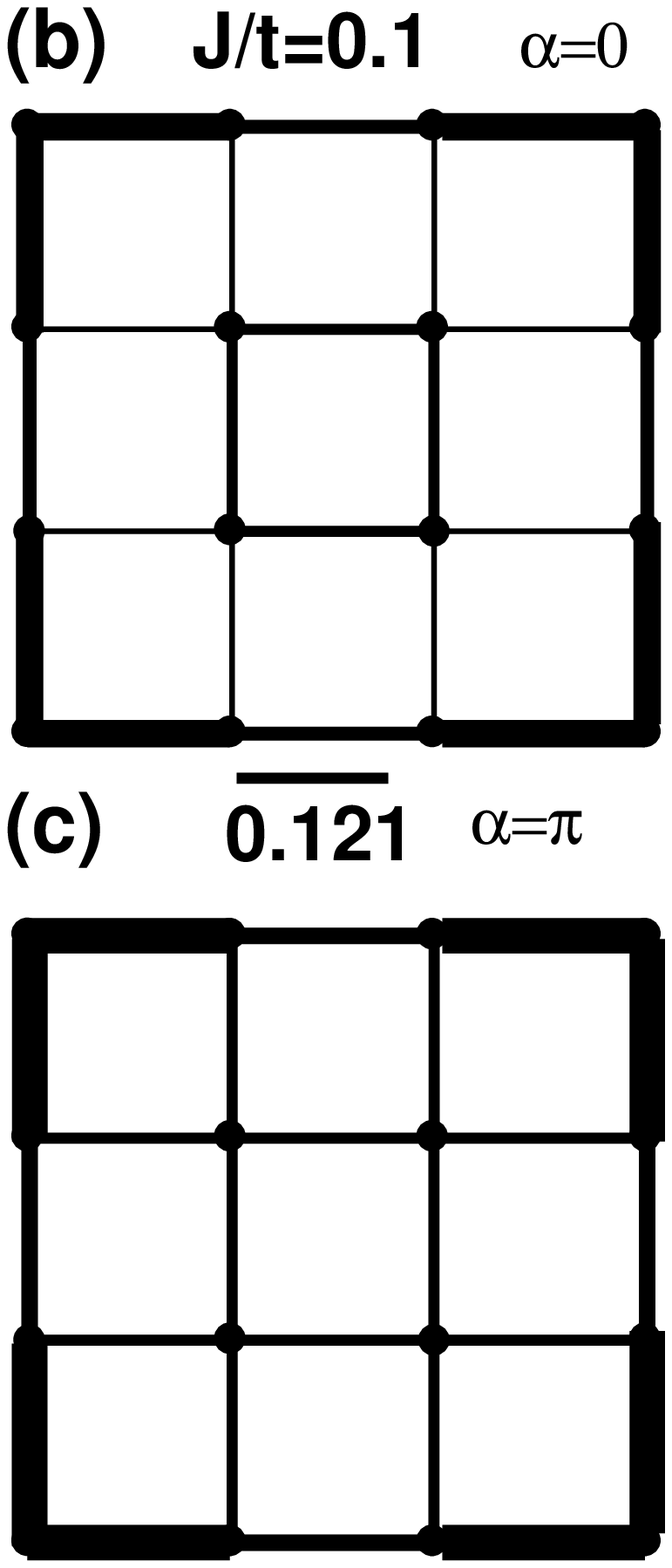}
\end{minipage}
\caption{(a) NN spin correlations in a $4 \times 4$ cluster with 2 holes for $J/t = 0.2$.
Despite some deviations for $\alpha=\pi /2$ and $7\pi /8$, the overall trend indicates that the
magnetic field promotes an increase of the NN AF correlations.
(b, c) Representation of the NN spin correlations in a $4\times4$ cluster 
(using OBC) with 2 holes for $J/t = 0.1$
at zero-field (b) and $\alpha=\pi$ (c). The thickness of the lines is proportional 
to the absolute value of the correlations. All values are negative (AF). 
The line on the top of figure 4c serves as a reference.
Note the overall increase of NN AF correlations with field. This shows that 
the results presented here do not depend qualitatively on the boundary conditions used.
}
\label{fig5}
\end{figure}

Figure 5a shows the variation with field of NN AF spin correlations in a $4\times4$ 
cluster with 2 holes, for $J/t = 0.2$, using PBC. Contrary to what was observed for 
the single-hole case, for 2 holes the increase of NN AF correlations
with magnetic field survives up to higher values of $J/t$ (up to $\approx 0.3$). This is
consistent with the idea that the field effect will be observable at higher 
$J/t$ values if the zero-field ground state is somehow weakened, since 
it is well known that in the $t-J$ model the short-range order associated with AF 
fluctuations is weakened at higher hole concentrations. At this point, one is then tempted to suggest, 
in view of the results for ladders, where the field seems to {\it reinforce} 
the SL character of the zero-field ground state, that also in the square 
clusters the field tries to create a SL state which has characteristics 
incompatible with those of the zero-field state. This sets up a competition 
which is regulated by doping and the value of $J/t$: higher dopings favor 
the field generated state and higher $J/t$ favors the zero-field state. 
One can now try to get a better idea of how this competition evolves 
by looking at spin correlations 
beyond NN and comparing the results with and without field. However, before doing that, 
it is important to verify how the results for square clusters depend 
on the boundary conditions. 
All the results showed up to this point have been for PBC. 
Figure 5b depicts NN correlations for $J/t=0.1$ in a $4 \times 4$ 
cluster with 2 holes at zero-field, now using
OBC. The thickness of the lines is proportional to the absolute value 
of the spin correlations between the connected sites
(all correlations shown are AF). For comparison, figure 5c shows the results at finite-field
($\alpha=\pi$).
It can be easily seen that the field enhances the NN AF spin correlations also 
for OBC, showing that the effect here presented is not dependent on boundary conditions.

\subsection{Effect of field in the longer range spin correlations}
\subsubsection{Review of zero-field results}

Before looking into the effect of the field in the longer range correlations, 
a discussion will be presented of how the zero-field static magnetic correlations 
at different 
dopings are described through ED calculations in a $4 \times 4$ cluster. 
It is our belief that the presentation of a somewhat detailed account of how the zero-field 
magnetic ground state in the $4 \times 4$ cluster evolves with doping from AF 
long-range-order (LRO) to AF SRO will clarify 
the description and interpretation of the finite-field results \cite{note-mag}. 
In figure 6a, correlations between spins at different distances are shown for a
$4 \times 4$ cluster at different
dopings at zero-field, for $J/t=0.1$ \cite{note-distances}.
Squares show spin correlations at zero-doping, and diamonds and circles display 
spin correlations for 1 and 2 holes, respectively.  It is well know, through 
calculations using different methods \cite{manousakis}, that the Heisenberg 
model in two dimensions (to which the $t-J$ model at zero-doping is equivalent)
presents AF LRO at zero temperature (2D N\'eel order). 
This {\it static} order translates, in a plot of spin correlations
as a function of distance between spins, into a finite-value plateau at long distances.
Because of the limited distances available
in a $4 \times 4$ cluster, the results in figure 6a for zero-doping (squares) only display a
{\it tendency} to level off and form a plateau \cite{plateau}. What would then be expected
as the $4 \times 4$ cluster is doped with the first hole? It is well
known from several experimental results \cite{kastner}, that the N\'eel phase (AF LRO) is
destroyed for a hole concentration $n_h \approx 0.02$. Therefore, for a doping
of $1/16$ ($n_h=0.0625$), one would naively 
expect the spin correlations in the $t-J$ model to display no trace of antiferromagnetism. 
However, the numerical results seem to indicate otherwise: judging from the curve for
$n_h=0.0625$ (diamonds) in figure 6a,
one would say that there are still quite robust AF tendencies in the doped system.
In reality, this is in agreement with experimental results, mainly neutron scattering 
\cite{kastner,birgeneau}, which describe this situation as AF SRO. 
In the cuprates, an AF exchange interaction between the copper oxide
planes leads to three-dimensional (3D) N\'eel order with a transition temperature
$T_N \approx 300 K$ (at zero-doping). It is this 3D order which quickly fades away 
at around $n_h=0.02$ and is represented by a narrow strip close to zero-doping 
in the phase diagram for the cuprates. However, strong AF correlations 
remain in the copper oxide plane. The energy spectra of these AF correlations (its fluctuations 
at different frequencies) has a very rich structure and a very complex dependence with doping. 
In addition, several aspects of the experimental results are not yet settled, different 
cuprate families presenting different results. However, due to the very 
strong AF exchange interaction between the copper spins in the planes, it is not 
totally surprising that AF scattering of some sort will be present in neutron experiments 
at dopings considerably higher than $n_h=0.02$. This surviving magnetism 
is generally referred to as AF SRO, and it is thought by many to play a very 
important role in defining the properties of
the cuprates for a wide region of the phase diagram. It 
can survive to moderate dopings, well into the metallic phase \cite{birgeneau}.
It is this AF SRO which is displayed by the somewhat robust 1 hole spin correlation 
results (diamonds) in figure 6a. 
Early ED results have shown that the $t-J$ model is able to capture this crucial 
property of the cuprates \cite{elbio-riera}.
When a second hole is introduced in the $4 \times 4$ cluster ($n_h=0.125$) the spin
correlations (circles) become ferromagnetic for the longest distance available
in the cluster (namely, $2 \sqrt{2}$) and are very close to zero for $\sqrt{5}$. This is 
again in good agreement with neutron scattering results, where the correlation length $\xi$ of
the AF SRO decreases with doping $n_h$ approximately as $\xi=3.8/\sqrt{n_h}$. Not coincidentally,
this formula also describes the variation with $n_h$ of the average separation between 
holes \cite{birgeneau}. At moderate doping, $\xi$ can be indeed very short:
in the metallic state of LSCO, for $n_h \approx 0.175$, one has that $\xi \approx 10 \AA$, 
which corresponds to approximately two lattice parameters. This is in qualitative agreement with 
the results for $n_h=0.125$ (circles) in figure 6a, since the distance 
where the correlations turn from AF to ferromagnetic provides an estimate of $\xi$. 
This discussion then shows that, despite the small size of the cluster analyzed, 
one can obtain reliable qualitative information, through the $t-J$ model, 
about some aspects of the magnetism in the ground state of the cuprates. 

\begin{figure}[h]
\centering
\includegraphics[width=10cm,angle=0]{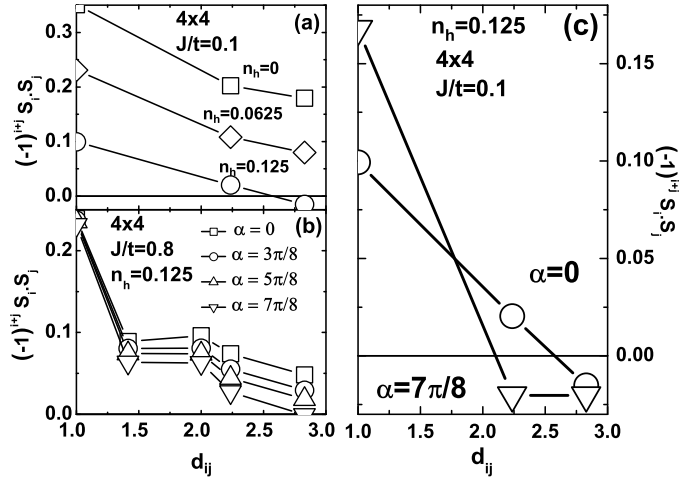}
\caption{ Variation of spin-spin correlations with distance for a $4 \times 4$ cluster.
Results are shown at zero-doping and with one and two holes, at zero- and finite-field.
(a) Results for $J/t=0.1$ at zero-field. Squares:
zero-doping, showing for a small square cluster how the AF LRO (2D N\'eel order)
is displayed. The tendency of the spin correlations to form a plateau at larger
distances is an indication of the AF LRO expected at the thermodynamic limit.
Diamonds: results for 1 hole ($n_h=0.0625$) indicate the transition from
N\'eel order to AF SRO at finite doping. Although the correlation
length is now finite, it is clearly larger than the cluster size.
Circles: results for 2 holes ($n_h=0.125$) indicate that the NN AF correlation is substantially
decreased if compared with the zero-doping case and that 
the correlations for larger distances are very close to zero (at a distance of $2\sqrt{2}$
the correlation becomes ferromagnetic). This indicates that the correlation length of
the AF SRO is of the order of the cluster size.
(b) Variation of the spin correlations with distance and field in a 
$4 \times 4$ cluster with 2 holes and $J/t=0.8$.
The decrease of the correlation length as the field increases can be clearly seen,
however the NN correlations are not affected (this is true for $J/t>0.3$).
(c) Comparison of spin correlations for $J/t=0.1$ with 2 holes at zero- and
finite-field. Notice that besides decreasing the correlation length, the field
also increases markedly the NN AF correlation, in contrast with results for $J/t>0.3$.
This suggests that at low values of $J/t$ the field induces a {\it qualitatively}
different ground state.
}
\label{fig6}
\end{figure}

\subsubsection{Field effect: Competition between AF SRO and SL?}

The discussion now turns to examining how {\it longer range spin correlations} are affected by 
the field. In the ladder results above, it was shown that spin correlations at 
larger distances than $\sqrt{2}$ were decreased by the external field \cite{sqrt2}. What are the 
results for square clusters? Although the authors are currently working on 
extensions and improvements of the present calculations \cite{newtjf}, for now 
a tentative picture can be summarized as follows. As mentioned already, the NN spin correlations increase 
substantially with field at low values of $J/t$, although this increase is suppressed 
for higher values of $J/t$, becoming negligible for $J/t > 0.3$, 
in a $4 \times 4$ cluster with 2 holes. However, non-negligible 
field effects are present, up to $J/t=2.0$, for all distances larger than NN. 
Above $J/t \approx 0.3$, the field effect at larger distances than NN is 
quite systematic, as can be seen in figure 6b for $J/t=0.8$, 
in a $4 \times 4$ cluster with 2 holes ($n_h=0.125$): 
the less AF the correlation is, i.e., the larger the distance between the spins, 
the larger is the decrease caused by the field, clearly showing the tendency of the field 
of trying to decrease the range of the AF SRO. However, it is important to notice that, 
for this value of $J/t$, the field 
effect in the NN correlation is virtually zero. Therefore, for values of $J/t$ where 
one would expect the AF SRO state to be quite robust, the field does not alter 
the qualitative aspects of the ground state. This picture remains essentially true down to 
$J/t \approx 0.3$. Below this value of $J/t$, there is a qualitative difference in 
the field effect, since there is a marked {\it increase} with field in the NN AF correlations. 
This is clearly seen in figure 6c, where results at zero- (circles) and finite-field 
(triangles) for $J/t=0.1$ with 2 holes in a $4 \times 4$ cluster clearly show the 
change from AF to ferro of the spin correlation at distance $\sqrt{5}$ and the 
pronounced increase of the NN AF correlation caused by the field.
The effect at larger distances can be still described as the field forcing a decrease in the 
correlation length, but the sharp increase in the NN AF correlation leads 
us to believe, motivated by the ladder results, that the finite-field ground state 
is {\it qualitatively different} from the zero-field ground state. The tentative use 
of the term {\it spin liquid} has its motivation more in the effect of the field 
in the NN correlations than in the longer range ones. The dependence of the results 
with hole doping and $J/t$ seems to be consistent with a SL-AF SRO competition. 
Therefore, this effect could be yet another example of
competition between different ground states relevant to the cuprates \cite{Zhang2}.

\begin{figure}[h]
\centering
\includegraphics[width=10cm,angle=0]{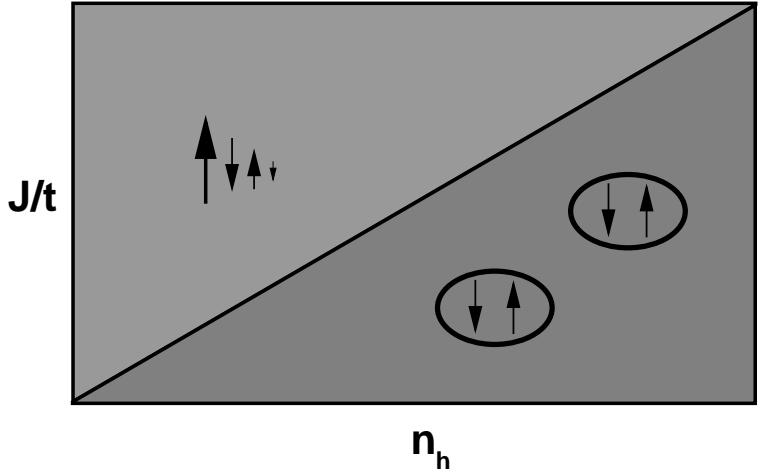}
\caption{Schematic representation of the phase diagram at 
finite-field. See text for details. 
Note that ideally, 
at the thermodynamic limit, the Nagaoka phase occurs only at $J/t=0$ 
and for vanishingly small hole concentration, therefore the phase 
boundary intercepts the origin.
}
\label{fig7}
\end{figure}

An schematic phase diagram at finite-field is displayed in figure 7. 
The main results obtained numerically for the 2D $t-J$ model in the presence 
of a perpendicular field are presented in a succinct way. In the left region, 
for high $J/t$ (vertical axis) and small $n_h$ (horizontal axis), the ground 
state presents AF SRO. Starting from a point inside 
the left-side region, as one increases $J/t$ and/or decreases $n_h$ (at 
constant field) the correlation length $\xi$ of the AF SRO increases (the 
system is drawn away from the phase boundary). On the other hand, 
if $J/t$ decreases and/or $n_h$ increases, a new ground state with 
SL characteristics is eventually established (the system crosses the phase boundary 
to the right-side region). An increase (decrease) of the field will rotate 
the phase boundary counter-clockwise (clockwise), enlarging 
the SL (AF SRO) phase and squeezing the AF SRO (SL) phase.

It is interesting to note that the effect observed in $4 \times 4$ clusters 
and 2-leg ladders is also present in 
a $2 \times 2$ plaquette with 1 and 2 holes. As expected, there are qualitative 
differences: the increase with field in NN AF correlations in a $2 \times 2$ plaquette is directly 
proportional to $J/t$ (therefore it vanishes at $J/t=0$ \cite{2x2}) and it is an order 
of magnitude smaller than the increase observed in ladders and $4 \times 4$ clusters.
Current efforts are underway to try to connect the results in $2 \times 2$ to the results described
in this paper and possibly gain a qualitative understanding of the magnetic field effect \cite{PRB}.

\section{Summary and conclusions}

In summary, ED calculations in 2-leg ladders and small square clusters have shown that applying a 
perpendicular magnetic field to the $t-J$ model at low doping tends to induce an SL-like state. 
In 2-leg ladders, the zero-field SL ground state is reinforced
by the field, independently of $J/t$ and doping.
Pair breaking caused by the field is clearly observed in ladders and its origin seems to be
associated to
the decrease with field in the across-the-diagonal AF spin correlation \cite{scalapino}.
On the other hand, in 2D this 
SL state has to compete with the zero-field AF SRO. The spin correlations of the true ground state
will then depend on the field strength, the value of $J/t$ and density of charge carriers. 
Interestingly enough, the field effect can be observed in clusters as small as a 
$2 \times 2$ plaquette. As the Hamiltonian in equation (1) can be diagonalized analytically 
in a $2 \times 2$ cluster, a qualitative 
understanding of the physical origin of the increase of the AF NN correlations may 
be achieved through a careful analysis of the eigenfunctions. One additional 
avenue of research worth investigating is how the results here presented will change 
if hoppings beyond NN are taken in account \cite{mart-pair}.

Because of the small size of the $4 \times 4$ cluster, the authors abstain of making any connection 
between the results here presented
and the experimental results mentioned in the introduction. However, given the relevance of the 
subject and the lack of calculations including the magnetic field in 
strongly correlated models like the $t-J$ and Hubbard, 
the authors firmly believe that the results so far presented will stimulate further investigations.
In particular, our results predict that an increase in the spin gap 
of doped ladders with applied field should be experimentally observable.

\ack
The authors wish to acknowledge helpful discussions with N. Bonesteel, W. Brenig, 
E. Dagotto, A. Moreo, A. Reyes, J. Riera 
and D. Veberi$\check{c}$.
G. M. acknowledges support from Martech (FSU), and A. F. A. from CNPQ-Brazil.


\begin{thebibliography}{<23>}
\bibitem{riso2} A comprehensive review of the neutron scattering results 
at finite magnetic field can be found in reference \cite{riso}
\bibitem{newtjf} Efforts are being made by the authors to extend the calculations 
to considerably larger square clusters.
\bibitem{flux} For one flux quantum threading each plaquette in the $4 \times 4$ cluster, the 
order of magnitude of the associated field is 1 KT.
\bibitem{note-bonds} It is interesting to note that the increase in the spin correlations of the 
rungs adjacent to the hole seems to be responsible for the substantial decrease 
in the spin correlations of the bonds in the leg opposite to the projected hole.
\bibitem{pair-note} The pair operators are defined as singlet pair operators in the rungs: 
$\Delta=(c_{i+}c_{j-}-c_{i-}c_{j+})$, where $i$ and $j$ are two sites in the same rung.
For details, see ref. \cite{mart-pair}
\bibitem{note-distances} For clarity of presentation, the figure concentrates in the 
NN correlations and the two largest distances in the $4 \times 4$ cluster. These 
are enough to present the main ideas in the discussion.
\bibitem{plateau} Quantum Monte Carlo (QMC) results for the Hubbard model 
at zero-doping in clusters with up to 64 sites 
can be seen in figure 14a of reference \cite{Elbio}. The very clear plateau presented there is 
an indication that the Hubbard model at zero-doping (as well as 
the $t-J$ model at zero-doping) captures the 
essential behavior of the spin background. Unfortunately, QMC results cannot be obtained at 
low temperatures for the doped $t-J$ or Hubbard models.
\bibitem{2x2} This is a revealing result. If the change in the correlations from 
ferromagnetic to AF at $J/t=0$ in the $4 \times 4$ cluster (Figure 4b) were a finite-size 
effect, then one would expect that it would be preserved as the size of the 
cluster {\it decreases}. Although this is not a proof, it suggests that 
the main effect discussed in this paper is not an artifact of the 
numerical method.
\end{thebibliography}
\end{document}